\documentclass[aps,twocolumn]{revtex4}
\usepackage[dvips]{graphics,graphicx}
\begin{document}
\title{Dipole and Bloch oscillations of cold atoms in a parabolic lattice}
\author{A.~V.~Ponomarev and A.~R.~Kolovsky} 
\affiliation{Kirensky Institute of Physics, Ru-660036 Krasnoyarsk, Russia}
\date{\today}

\begin{abstract}
The paper studies the dynamics of a Bose-Einstein condensate
loaded into a 1D parabolic optical lattice, and excited by a sudden
shift of the lattice center. Depending on the magnitude of the initial shift, 
the condensate undergoes either dipole or Bloch oscillations. 
The effects of dephasing and of atom-atom interactions on
these oscillations are discussed.
\end{abstract}

\maketitle

1. Bloch oscillations (BO) of a quantum particle in a periodic
potential are one of the most fascinating phenomena of quantum
physics \cite{Zene34}. Since the pioneering experiment \cite{Daha96}
in 1996, this phenomenon has been intensively studied
for cold atoms in optical lattices \cite{63}, with recent
emphasis on quantum statistical (Fermi or Bose)
and atom-atom interaction effects. In particular, the dynamics of 
degenerate Bose gases, on which we will focus here, was studied
experimentally in \cite{Mors01,Scot04,Modu04}. It should
be stressed from the very beginning that, when addressing this
problem theoretically, one has to distinguish between quasi
one-dimensional lattices (created by two counter-propagating laser
beams) and truly 1D lattices (or so-called modulated quantum
tubes). Indeed, in the former case the number of atoms per well
of the optical lattice can be as large as $10^3-10^4$, and a
mean field approach (based on the Gross-Pitaevskii or nonlinear
Schr\"odinger equation) is generally justified. This is not the
case of the truly 1D lattices, where only few atoms occupy a single
well, and, hence, a microscopic analysis is required. For a
tilted infinite lattice such analysis, based on the
Bose-Hubbard model, was presented in 
\cite{57,61,60}, where two regimes of BO -- quasiperiodic and
irreversible decaying -- were identified.

When referring to the typical laboratory experiments, an additional
complication stems from the harmonic confinement along the lattice.
Clearly, harmonic confinement should modify BO of bosonic atoms, and
the aim of this work is to estimate its effect. At the same
time, parabolic lattices have their own interest, because
they allow to study dipole oscillations of BECs. Recent
experiments \cite{Fert05} have shown that there is a fundamental
difference between dipole oscillations in quasi-
and truly 1D lattices. While in the former case the main effect of
the periodic potential can be taken into account by simply
substituting the atomic mass by its effective mass in the ground
Bloch band \cite{Cata03}, one observes a rapid decay of
oscillations in the latter case. In the present paper we also
briefly discuss dipole oscillations of a BEC in truly 1D
lattices, partially overlapping in this part with recent
theoretical work \cite{Rey05}.

2. We consider atoms in a parabolic lattice potential
$V(x)=M\omega^2x^2/2-V\cos^2(2\pi x/d)$, where the atoms are set
into motion by a sudden shift of the trap origin. The relevant
parameters of the system are the hopping matrix element $J$
(defined by the amplitude of the periodic potential $V$),
the `parabolicity' $\nu=M\omega^2 d^2$, and the initial shift $l_0=\Delta x/d$.
Using the single band approximation and neglecting atom-atom interactions, the
dynamics of the system is described by the pendulum model \cite{63,preprint},
\begin{equation}
\label{1}
i\hbar\dot{a}_l=\frac{\nu}{2}l^2a_l
-\frac{J}{2}\left(a_{l+1}+a_{l-1}\right) \;,
\end{equation}
where $a_l(t)$ is the complex amplitude of the atoms in the $l$-th
well of the optical lattice. The separatrix of the pendulum corresponds to the
shift 
\begin{equation}
\label{2}
l^*=2(J/\nu)^{1/2} \;.
\end{equation}
If the initial shift $l_0<l^*$, the pendulum shows oscillations around the
equilibrium point and, referring to the original system, this regime is
regarded as that of dipole oscillations of the atoms. If $l_0>l^*$,
the pendulum is in the rotational regime, and the dynamics of the
atoms can be regarded as BO in a local static field $F=\nu l_0/d$.

3. We begin with analyzing the regime of BO for an ideal bosonic
system (i.e., no atom-atom interactions).
Because the local static force $F$ is not homogeneous, one has an
additional process of dephasing of BO in a parabolic lattice, as
compared to the paradigmatic case of a homogeneously tilted lattice.
When discussing the mean atomic momentum, we can estimate this
effect of dephasing by evaluating the sum
\begin{equation}
\label{3}
p(t)\sim\sum_m \exp(-m^2/2\gamma^2) \sin[(\omega_B+\nu m/\hbar)t] \;,
\end{equation}
where $\omega_B=\nu l_0/\hbar$, $m=l-l_0$, and $\gamma$ is the width of the atomic wave packet (measured in units of the lattice period). Substituting the sum by an integral, we obtain
\begin{equation}
\label{4}
p(t)\sim\exp(-t^2/2\tau_\gamma^2)\sin(\omega_B t) \;,
\end{equation}
where
\begin{equation}
\label{5}
\tau_\gamma=\hbar/\gamma\nu \;.
\end{equation}
It is seen in Eq.~(\ref{5}),  that the dephasing time $\tau_\gamma$
is defined by both the wave packet width and by the trap frequency
(see Fig.~\ref{fig1} below). On the basis of this result one might
conclude that a narrow wave packet is preferable for studying BO
in parabolic lattices. This is, however, not exactly true
because a narrow wave packet implies a lower contrast of the
interference pattern measured in laboratory experiments. Thus,
one has to keep a compromise between the contrast and dephasing,
when preparing the initial wave packet.

The irreversible decay of BO according to Eq.~(\ref{4}) is a
consequence of our approximation of the sum by an integral.
Without this approximation, the decay of oscillations is followed
by periodic revivals with a period $T_\nu=2\pi\hbar/\nu$
\cite{remark1}. One of these revivals is illustrated in the upper
panel of Fig.~\ref{fig1}, which shows the dynamics of the mean
momentum of the non-interacting atoms in the parabolic lattice
with parabolicity $\nu=0.04 J$.  As initial state of the
system we choose here the ground state of the atoms in a
parabolic lattice with a slightly tighter confinement $\nu'=4\nu$,
which was then shifted by a distance $l_0=8l^*=80$. Note that
by changing $\nu'$ we change only the dephasing time [through the
change of the wave packet width $\gamma=\gamma(\nu')$], while the
revival time is defined exclusively by the parameter $\nu$.
\begin{figure}[t!]
\center
\includegraphics[width=8.5cm, clip]{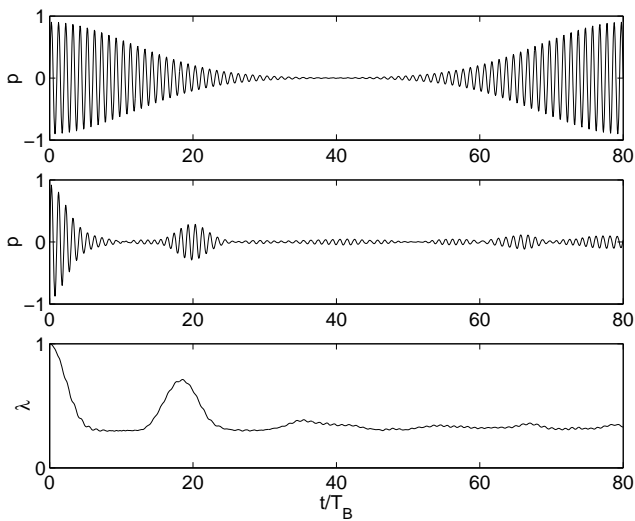}
\caption{Bloch oscillations of $N=5$ atoms in a parabolic
lattice with parabolicity $\nu=0.04J$: (a) mean momentum of
non-interacting atoms; (b) mean momentum of interacting
($W=0.2J$) atoms; (c) macroscopic coherence of the 
interacting atoms. Initial shift $l_0=8l^*=80$.} \label{fig1}
\end{figure}

4. Next we address the effect of atom-atom interactions. In the
case of quasi one-dimensional parabolic lattices, Bloch and dipole
oscillations of the interacting atoms were studied in a number of
papers, using the mean-field approach
\cite{Kram02,Chio01,Smer02,Adhi03,Nesi04,preprint}. As known, the
mean-field approach is justified in the limit of large occupation
number $\bar{n}\rightarrow\infty$ and vanishing microscopic interaction
constant $W\rightarrow 0$,  and leads (in the simplest case) to the
discrete nonlinear Schr\"odinger equation,
\begin{equation}
\label{6}
i\hbar\dot{a}_l=\frac{\nu}{2}l^2 a_l
-\frac{J}{2}\left(a_{l+1} +a_{l-1}\right)+g|a_l|^2a_l \;,
\end{equation}
where $g=WN$ is the macroscopic interaction constant. In our
present work, we focus on the case of truly one-dimensional
lattices, where the mean occupation number $\bar{n}\sim 1$.
Clearly, the mean field approach is not applicable here and one
has to treat the system microscopically by using, for example, the
Bose-Hubbard model,
\begin{equation}
\label{7}
H=\sum_l \frac{\nu}{2} l^2 \hat{n}_l
-\frac{J}{2}\left(\sum_l \hat{a}^\dag_{l+1}\hat{a}_l+h.c.\right)
  +\frac{W}{2}\sum_l \hat{n}_l(\hat{n_l}-1) \;.
\end{equation}
The main question we address below is the effect of atom-atom
interactions on the Bloch dynamics depicted in the upper panel of
Fig.~\ref{fig1}.

First we shall discuss the initial conditions in some more detail.
Throughout the paper we shall consider the ground many-body state
of the atoms in a parabolic lattice as the initial wave packet
(which is shifted then by the distance $l_0$). Clearly, along with
the ratio $J/\nu$ this state is also defined by the ratio of the
interaction constant to the hopping matrix element. Namely, it is
essentially given by the symmetrized product of the
single-particle atomic state for $W<J$ , while it may resemble
the Mott-insulator state for $W\gg J$ \cite{Rey05}. In what follows we restrict
ourselves by a relatively weak interaction. Then the ground state
of the system can be well approximated by the many-body wave function
\begin{equation}
\label{9}
|\tilde{\Psi}_0\rangle=\sum_{\bf n} c_{\bf n}|\bf{n}\rangle \;,\quad
c_{\bf n}=\sqrt{N!}\prod_l\frac{a_l^{n_l}}{\sqrt{n_l!}} \;,
\end{equation}
where $|{\bf n}\rangle=|\ldots,n_{-1},n_0,n_1,\ldots\rangle$ is
the Fock basis and the $a_l$ satisfy the stationary nonlinear
Schr\"odinger equation
\begin{equation}
\label{10}
\frac{\nu}{2}l^2 a_l-\frac{J}{2}\left(a_{l+1}+a_{l-1}\right)
+g|a_l|^2a_l=E_0a_l \;.
\end{equation}
For example, for $N=5$, $\nu=0.04J$ and $W=0.2J$, the overlap of
the state (\ref{9}) with the exact ground state $|\Psi_0\rangle$
is $|\langle\Psi_0|\tilde{\Psi}_0\rangle|^2=0.97$. We note that
the state (\ref{9}) is completely coherent and is analogous to the
super-fluid state in a homogeneous lattice. We shall characterize
the macroscopic coherence of the given many-body state
$|\Psi\rangle$ by the maximal eigenvalue $\lambda$ of the
single-particle density matrix
\begin{equation}
\label{8}
\rho_{l,m}=N^{-1}\langle\Psi|\hat{a}_l^+\hat{a}_m|\Psi\rangle \;.
\end{equation}
Then the macroscopic coherence of the state (\ref{9}) is $\lambda=1$.

We proceed with the dynamics. The middle panel in Fig.~\ref{fig1}
shows the mean momentum of $N=5$ interacting atoms ($W=0.2J$). In
comparison with the noninteracting case (upper panel), a qualitative
change is noticed. This change can be understood by analyzing the
macroscopic coherence of the system, shown in the lower panel. It
is seen that the macroscopic coherence oscillates with some
characteristic period $T_W$. In the case of a homogeneously tilted
lattice these oscillations were studied in Ref.~\cite{57}. The origin
of the oscillations was shown to be the Stark localization of the
single-particle wave functions which, together with the discreetness
of the atom number, leads to the following expression for the
macroscopic coherence,
\begin{equation}
\label{11} \lambda=\exp(-2\bar{n}[1-\cos(Wt/\hbar)]) \;.
\end{equation}
In Eq.~(\ref{11}) $\bar{n}$ is the mean number of atoms per
lattice site \cite{remark2} and the limit $Fd\gg J$ is implicitly
assumed. Since for the considered local static force $Fd=\nu
l_0=3.2J$ Stark localization is not complete, the oscillations
of the macroscopic coherence decay in time.
Nevertheless, if this irreversible decay of coherence is slow on
the time scale of the dephasing time, one can observe the revival
of BO of the interacting atoms -- an effect which attracts much
attention because it provides an independent and accurate method
for measuring the microscopic interaction constant $W$.

\begin{figure}[t!]
\center
\includegraphics[width=8.5cm, clip]{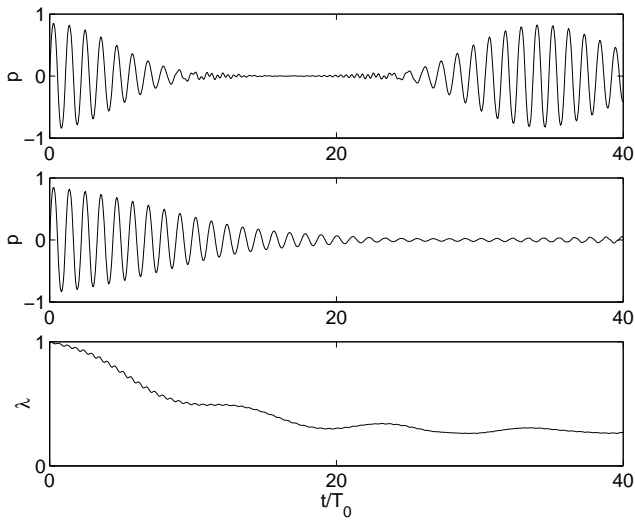}
\caption{Dipole oscillations of $N=4$ atoms in the parabolic
lattice with parabolicity $\nu=0.04J$: (a) mean momentum of
non-interacting atoms; (b) mean momentum of interacting
($W=0.2J$) atoms; (c) macroscopic coherence of interacting
atoms. Initial shift $l_0=l^*/2=5$.} \label{fig5}
\end{figure}

5. Let us now turn to the case $l_0<l^*$. Here we meet dipole
oscillations of a BEC with a characteristic frequency given by the
frequency of small pendulum oscillations $\omega_0=(\nu
J)^{1/2}/\hbar$. (We recall in passing that the frequency of BO
was given by $\omega_B=\nu l_0/\hbar\approx 2\omega_0 l_0/l^*$,
$l_0\gg l^*$.) For vanishing atom-atom interactions these dipole
oscillations are shown in the upper panel of Fig.~\ref{fig5},
where $l_0=l^*/2=5$, and the other parameters are the same as in
Fig.~\ref{fig1}. The dephasing time $\tau_\gamma$ is again given
by Eq.~(\ref{5}) but with the parameter $\nu$ substituted by the
nonlinearity parameter $\tilde{\nu}=\nu/8$ \cite{Lich83}. (The
latter parameter also defines the revival time.) The middle and
lower panels in Fig.~\ref{fig5} refer to interacting atoms.
An exponential decay of the macroscopic coherence is noticed. The
other point to which we want to draw the attention of the reader is that a
moderate interaction stabilizes the dipole oscillations against
dephasing. Within the mean-field approach (which reduces the
Bose-Hubbard model to the discrete nonlinear Schr\"odinger
equation), this phenomenon is discussed in Ref.~\cite{preprint}.

6. In conclusion, we have shown that the dynamics of cold atoms in
parabolic lattices is governed by the relation between two
characteristic times -- the dephasing time $\tau_\gamma$ and the
decoherence time $\tau_W$.

The dephasing time is inversely proportional to the width $\gamma$
of the initial wave packet and the nonlinearity $\tilde{\nu}$,
which, in turn, is defined by the initial shift $l_0$ of the wave
packet relative to the separatrix $l^*=2(J/\nu)^{1/2}$. Namely,
$\tilde{\nu}=\nu/8$ for $l_0\ll l^*$, and $\tilde{\nu}=\nu$ for
$l_0\gg l^*$. It is interesting to estimate the dephasing
time in a typical laboratory experiment. Taking, as an example,
the recent experiment \cite{Fert05} with rubidium atoms in an
array of axially modulated quantum tubes, we have $\nu=0.0014E_R$
and $J=0.38E_R$ for the modulation amplitude (depth of the optical
lattice) of one recoil energy. This gives a separatrix position $l^*=33$,
and a period $T_0=12.1\ \rm ms$ of small dipole oscillations. Assuming
a dilute gas (which, in fact, is not the case realized in the
cited experiment) the width of the initial wave packet is
$\gamma\approx(J/4\nu)^{1/4}=\sqrt{l^*}/2\approx3$ and, hence, the
dephasing time $\tau_\gamma=85\ \rm ms$ for dipole oscillations, and
$\tau_\gamma=10.6\ \rm ms$ for BO. Note that these are upper
estimates for the dephasing times, and for the initial shift $l_0$
closer to the separtrix the dephasing times are essentially
smaller. It is also worth noting that there is a maximal shift
$l_0$ above which the single band approximation (used throughout
the paper) is not valid. The crucial parameter here is the
energy gap between the Bloch bands ($\Delta=0.5E_R$ for the
specified parameters). The analysis of BO in a parabolic lattice
beyond the single-band approximation will be subject of a
separate paper.

The decoherence time $\tau_W$ is defined by the characteristic
density of the atomic gas $\bar{n}$ and by the value of the
microscopic interaction constant $W$. The latter, in turn, is
defined by the $s$-wave scattering length and by the degree of
confinement of the atoms in the wells of the optical potential. In
particular, in the experiment \cite{Fert05}, the quantum tubes
were created by two crossing quasi 1D optical lattices with an
amplitude $V=30E_R$. For the axial modulation with $V=E_R$ this
gives $W=0.73E_R$. For this relatively high value of the
interaction constant few atoms per one tube are enough to destroy
the dipole/Bloch oscillations on a very short time scale. This
qualitatively explains the results of the experiment
\cite{Fert05}, where the number of atoms per one quantum tube was
around 20. To observe the effects discussed in this paper one has
to decrease either the atomic density or the interaction constant
(e.g. by use of a Fishbach resonance), as compared to those of
Ref.~\cite{Fert05}.
%



\begin{thebibliography}{10}

\bibitem{Zene34}
C.~Zener, Proc. R. Soc. Lond. {\bf A145}, 523  (1934).

\bibitem{Daha96}
M.~Ben Dahan, E.~Peik, J.~Reichel, Y.Castin, and C.~Salomon,
Phys. Rev. Lett. \textbf{76}, 4508 (1996).

\bibitem{63}
For the introductory review see A.R.Kolovsky and H.J.Korsch,
International J. of Mod. Physics {\bf 18}, 1235 (2004).

\bibitem{Mors01}
O.~Morsch, J.~H.~M\"uller, M.~Cristani, D.~Ciampini, and E.~Arimondo,
Phys. Rev. Lett. \textbf{87}, 140402 (2001).

\bibitem{Scot04}
R.~G.~Scott, A.~M.~Martin, S.~Bujkiewicz, T.~M.~Fromhold,
N.~Malossi, O.~Morsch, M.~Cristiani, and E.~Arimondo,
Phys. Rev. A {\bf 69}, 033605 (2004).

\bibitem{Modu04}
M.~Modugno, E.~de~Mirandes, F.~Ferlaino, H.~Ott, G.~Roati, and M.~Inguscio,
Fortschr. Phys. \textbf{52}, 1173 (2004).

\bibitem{57}
A.~R.~Kolovsky,
Phys. Rev. Lett. \textbf{90}, 213002 (2003).

\bibitem{61}
A.~Buchleitner and A.~R.~Kolovsky,
Phys. Rev. Lett. \textbf{91}, 253002 (2003).

\bibitem{60}
A.~R.~Kolovsky and A.~Buchleitner,
Phys. Rev. E {\bf 68}, 056213 (2003).

\bibitem{Fert05}
C.~D.~Fertig, K.~M.~O'Hara, J.~H.~Huckans, S.~L.~Rolston, W.~D.~Phillips,
and J.~V.~Porto,
Phys. Rev. Lett. {\bf 94}, 120403 (2005).

\bibitem{Cata03}
F.~S.~Cataliotti, L.~Fallani, F.~Ferlaino, C.~Fort, P.~Maddaloni, and M.~Inguscio,
New J. of Phys. \textbf{5}, 71.1 (2003).

\bibitem{Rey05}
A.~M.~Rey, G.~Pupillo, C.~W.~Clark, and C.~J.~Williams,
e-print: cond-mat/0503477.

\bibitem{preprint}
J.~Brand and A.~R.~Kolovsky,
e-print: cond-mat/0412549.

\bibitem{remark1}
Indeed, at times $t$ multiple to $T_\nu$ the phases $\nu mt/\hbar$ in
the sum (\ref{3})  are multiple of $2\pi$ and, hence, the momentum takes
its initial value. (More rigourosly, the revivals of ocsillations follow
from the quadratic dependence for the eigenenergies of the quantum pendulum
in the asymptotic region $l\gg l^*$.)

\bibitem{Kram02}
M.~Kr\"amer, L.~Pitaevskii, and S.~Stringari,
Phys. Rev. Lett. \textbf{88}, 180404 (2002).

\bibitem{Chio01}
M.~L.~Chiofalo and M.~P.~Tosi,
J. Phys. B: At. Mol. Opt. Phys. {\bf 34}, 4551 (2001).

\bibitem{Smer02}
A.~Smerzi, A.~Trombettoni, P.~G.~Kevrekidis, and A.~R.~Bishop,
Phys. Rev. Lett. \textbf{89}, 170402 (2002).

\bibitem{Adhi03}
S.~K.~Adhikari,
Eur. Phys. J. D {\bf 25}, 161 (2003).

\bibitem{Nesi04}
F.~Nesi and M.~Modugno,
J.~Phys.~B: At. Mol. Opt. Phys. {\bf 37}, S101 (2004).

\bibitem{remark2}
In the considered case of a finite wave packet it looks
reasonable to associate $\bar{n}$ with the number of atoms $N$
devided by the wave packet width $\gamma$.

\bibitem{Lich83}
A.~J.~Lichtenberg and M.~A.~Libermann, {\em Regular and chaotic
dynamics} (Springer, Berlin, 1983).


\end{thebibliography}
\end{document}